# Extension of the critical inclination

Xiaodong Liu[1], Hexi Baoyin[2], and Xingrui Ma[3]

*School of Aerospace, Tsinghua University, Beijing 100084 CHINA*

Email:   *liu-xd08@mails.tsinghua.edu.cn*; *baoyin@tsinghua.edu.cn*; *maxr@spacechina.com*

**Abstract**

The critical inclination is of special interest in artificial satellite theory. The critical inclination can maintain minimal deviations of eccentricity and argument of pericentre from the initial values, and orbits at this inclination have been applied to some space missions. Most previous researches about the critical inclination were made under the assumption that the oblateness term $J_2$ is dominant among the harmonic coefficients. This paper investigates the extension of the critical inclination where the concept of the critical inclination is different from that of the traditional sense. First, the study takes the case of Venus for instance, and provides some preliminary results. Then for general cases, given the values of argument of pericentre and eccentricity, the relationship between the multiplicity of the solutions for the critical inclination and the values of $J_2$ and $J_4$ is analyzed. Besides, when given certain values of $J_2$ and $J_4$, the relationship between the multiplicity of the solutions for the critical inclination and the values of semimajor axis and eccentricity is studied. The results show that for some cases, the value of the critical inclination is far away from that of the traditional sense or even has multiple solutions. The analysis in this paper could be used as starters of correction methods in the full gravity field of celestial bodies.

---

[1]  PhD candidate, School of Aerospace, Tsinghua University

[2]  Associate Professor, School of Aerospace, Tsinghua University

[3]   Professor, School of Aerospace, Tsinghua University

## 1. Introduction

In the theory of artificial satellites, the critical inclination is always a focus of researches. The concept of the critical inclination was first introduced by Orlov (Orlov 1953). In order to deal with the orbits at the critical inclination, the short-periodic terms were eliminated based on canonical transformations (Brouwer 1958). By making use of numerical integrations, geometrical interpretations of the critical inclination were provided (Coffey et al. 1986). Many other early researches were contributed to the problem of the critical inclination, the details of which can be seen in (Jupp 1988).

As known in celestial mechanics, orbits at the critical inclination take the critical inclination to keep eccentricity and argument of perigee invariable on average. For Earth satellites, the Molniya (Stone and Brodsky 1988; Kidder and Vonder Haar 1990; Gunning and Chao 1996) and Tundra orbits (Barker and Stoen 2001; Bruno and Pernicka 2002; Bruno and Pernicka 2005) applied such conditions to stop the rotation of argument of pericentre and the variation in eccentricity. Orbits around the Moon were also studied in order to reduce the need for stationkeeping (Delhaise and Morbidelli 1993; Ely and Lieb 2005; Saedeleer and Henrard 2006). Some researches have regarded orbits at the critical inclination as types of frozen orbits (Coeefy et al. 1994; Lara et al. 1995; Aorpimai and Palmer 2003; Russell and Lara 2007; Liu et al. 2011).

Most previous researches about the critical inclination were made under the assumption that the oblateness term $J_2$ is dominant among the harmonic coefficients. This assumption is effective for most large celestial bodies, including Earth, Mars, and Moon. However, there exist some celestial bodies where the other first few harmonic coefficients are of the same order of magnitude as the oblateness term $J_2$, or even greater than $J_2$. For example, the $J_3$ and $J_4$ terms of Venus are of the same order of magnitude as $J_2$. For these central bodies, the concept of the critical inclination is different from that of the traditional sense. In the present paper, the extended problem of the critical inclination is considered. It is found that for some cases, the value of the critical inclination is far away from that of the traditional sense or even has multiple solutions. The investigations of the extension of the critical inclination could provide good initial conditions for numerical correction methods in the more complex models of celestial bodies.

**2. Critical inclination in the traditional sense**

As known in celestial mechanics, both argument of pericentre and eccentricity can remain constant on average at the critical inclination. According to the first order theory, the secular perturbations of the spacecraft only include the effect of the oblateness term $J_2$. Then the averaged variational rates of argument of pericentre and eccentricity are (Chobotov 2002)

$$\dot{\omega}_1 = -\frac{3nJ_2R_e^2}{2a^2(1-e^2)^2}\left(\frac{5}{2}\sin^2 i - 2\right) \quad (1)$$

$$\dot{e}_1 = 0 \quad (2)$$

where $a$ is semimajor axis; $e$ is eccentricity; $i$ is inclination; $\omega$ is argument of pericentre; $J_2$ is the zonal harmonic of the second order (also known as the oblatenes term); $R_e$ is the reference radius of the central body; $n$ is the mean angular velocity, and $n = \sqrt{\mu/a^3}$.

The averaged variational rate of eccentricity is always equal to zero. It is evident that the variation in eccentricity can be stopped if the inclination yields

$$\frac{5}{2}\sin^2 i - 2 = 0 \quad (3)$$

Then the critical inclination of the traditional sense can be easily obtained

$$i_{c0} = \cos^{-1}\left(1/\sqrt{5}\right) \cong 63.4349° \,(\text{or}\, 116.5651°) \quad (4)$$

Thus, orbits at the inclinations in the neighborhood of the critical inclination $i_{c0}$ are effectively frozen when only considering the first order perturbation involving with the oblateness term $J_2$.

## 3. A case of Venus

The criterion for the critical inclination in Section 2 is effective if the oblateness term $J_2$ is dominant among the harmonic coefficients. However, this criterion fails to converge when the oblateness term $J_2$ is not dominant among the harmonics. For this case, the value of the critical inclination may be far away from $i_{c0}$ or even have multiple solutions.

Table 1 lists the zonal harmonics up to degree 4 for four different celestial bodies: Earth (Lemoine et al. 1998), Mars (Lemoine et al. 2001), Moon (Konopliv et al. 2001) and Venus (Konopliv et al. 1999). It can be seen that the gravity field of Venus is quite different from that of other celestial bodies. For Earth, the other first few harmonic coefficients are about 3 orders of magnitude lower than $J_2$; for Mars, the other first few harmonic coefficients are about 2 orders of magnitude lower than $J_2$; for Moon, the other first few harmonic coefficients are about 1–2 orders of magnitude lower than $J_2$; while for Venus, the other first few harmonic coefficients are of the same order of magnitude as $J_2$. Thus, the effect of the other terms of the harmonic coefficients cannot be neglected for Venus, so the criterion for the critical inclination in Section 2 is no longer effective.

**Table 1** Harmonic coefficients for Earth, Mars, Moon and Venus.

| Harmonics | Earth | Mars | Moon | Venus |
|---|---|---|---|---|
| $J_2$ | $1.0826 \times 10^{-3}$ | $1.9555 \times 10^{-3}$ | $2.0323 \times 10^{-4}$ | $4.4044 \times 10^{-6}$ |
| $J_3$ | $-2.5327 \times 10^{-6}$ | $3.1450 \times 10^{-5}$ | $8.4759 \times 10^{-6}$ | $-2.1082 \times 10^{-6}$ |
| $J_4$ | $-1.6196 \times 10^{-6}$ | $-1.5377 \times 10^{-5}$ | $-9.5919 \times 10^{-6}$ | $-2.1474 \times 10^{-6}$ |

According to mean element theory, the secular perturbations of the first order and second order due to the gravitational asphericity depend on the $J_2$ and $J_4$ terms, while the $J_3$ term gives rise only to short periodic and long periodic effects (Brouwer 1959; Kozai 1959). In celestial mechanics, the mean motion of the spacecraft is always of special interest. The motion of a satellite in the potential field involving with $J_2$ and $J_4$ terms was often studied for theoretical analysis (Garfinkel 1960; Izsak 1962; Allan 1970; Garfinkel 1973). Therefore, in this paper, only the perturbations of the $J_2$ and $J_4$ terms are considered. The gravitational potential of a satellite in the $J_2$ and $J_4$ gravity field can be expressed as

$$U = \frac{\mu}{r}\left[1 - J_2\left(\frac{R_e}{r}\right)^2 P_2(\sin\varphi) - J_4\left(\frac{R_e}{r}\right)^4 P_4(\sin\varphi)\right] \tag{5}$$

where $\mu$ is the gravitational constant; $r$ is the distance of the spacecraft; $P_l$ is the Legendre function of degree $l$; $\varphi$ is the latitude of body-fixed coordinate system.

The averaged variational rates of eccentricity and argument of pericentre due to the secular perturbations of the first order are shown in Eqs. (1) and (2), and the averaged variational rates of argument of pericentre and eccentricity due to the secular perturbations of the second order are represented as (Brouwer 1959)

$$\begin{aligned}\dot{\omega}_2 = \frac{9nJ_2^2 R_e^4}{4a^4(1-e^2)^4}&\left\{\left(4 + \frac{7}{12}e^2 + 2\sqrt{1-e^2}\right) - \sin^2 i\left(\frac{103}{12} + \frac{3}{8}e^2 + \frac{11}{2}\sqrt{1-e^2}\right)\right.\\ &+ \sin^4 i\left(\frac{215}{48} - \frac{15}{32}e^2 + \frac{15}{4}\sqrt{1-e^2}\right)\\ &\left. - \frac{35J_4}{18J_2^2}\left[\left(\frac{12}{7} + \frac{27}{14}e^2\right) - \sin^2 i\left(\frac{93}{14} + \frac{27}{4}e^2\right) + \sin^4 i\left(\frac{21}{4} + \frac{81}{16}e^2\right)\right]\right\}\end{aligned} \tag{6}$$

$$\dot{e}_2 = 0 \tag{7}$$

Thus, the averaged variational rates of eccentricity of eccentricity and argument of pericentre are written in the form

$$\dot{\bar{\omega}} = \dot{\omega}_1 + \dot{\omega}_2 = 0 \tag{8}$$

$$\dot{\bar{e}} = \dot{e}_1 + \dot{e}_2 = 0 \tag{9}$$

Eq. (9) is naturally met. Seen from Eq. (8), the corresponding critical inclination can be obtained if setting the values of semimajor axis $a$ and eccentricity $e$, and it is no longer a constant.

As the first step, the case of Venus is taken as an example. Given the semimajor axis $a=7000$ km and eccentricity $e=0.1$, the corresponding critical inclination is $i_c=52.5825°$, which is far away from the value of $i_{c0}$. Figure 1(a) shows the evolution of $e$ and $\omega$ over 100 days in the zonal gravity field up to degree 4 for the critical inclination $i_c=52.5825°$. It can be seen that the $e-\omega$ evolution is almost a point in Fig. 1(a), so the drift rates of $e$ and $\omega$ are approximately equal to zero. Figure 1(b) presents the evolution of $e$ and $\omega$ over 100 days in the zonal gravity field up to degree 4 for $i=20°$ that is far away from the critical inclination. It can be seen that $\omega$ keeps increasing during 100 days, the magnitude of which is approximately $7°$, and $e$ also keeps increasing, the magnitude of which is approximately 0.005. In Section 4, it is found that the Venusian $J_2$ and $J_4$ lie on the region of one solution for the critical inclination.

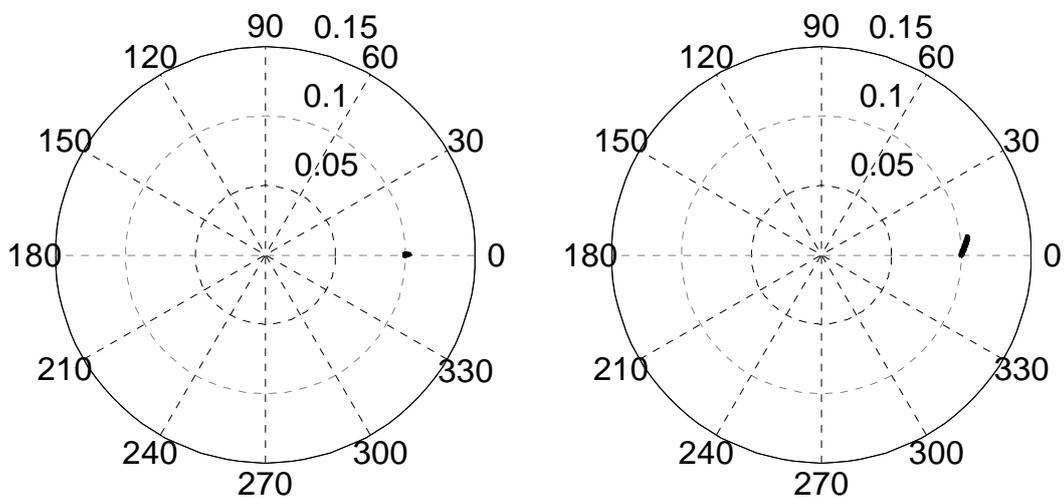

(a)　　　　　　　　　　　　　　(b)

**Fig. 1** $e - \omega$ evolution over 100 days in the zonal gravity field up to degree 4. (**a**) $i_c$=52.5825°. (**b**) $i$=20°.

## 4. General Cases

From Section 3, it can be seen that for Venus, the value of the critical inclination depends on semimajor axis and eccentricity, and may be far away from that of $i_{c0}$ in the traditional sense. In this section, two different cases are investigated. First, when given the values of argument of pericentre and eccentricity, the relationship between the multiplicity of the solutions for the critical inclination and the values of $J_2$ and $J_4$ is analyzed. Second, given certain values of $J_2$ and $J_4$, the relationship between the multiplicity of the solutions for the critical inclination and the values of semimajor axis and eccentricity is studied.

In order to make the analyses easier and clear, the scaling is made that the reference radius $R_e$ is the unit of length, the mass of the central body $M$ is the unit of mass, and $t_c$ is the unit of time, where

$$t_c = \sqrt{\frac{R_e^3}{GM}} \tag{10}$$

Eq. (8) can be written in a more concrete form

$$f(\sin i) = A \sin^4 i + B \sin^2 i + C = 0 \tag{11}$$

where

$$A = \frac{9nJ_2^2 R_e^4}{4p^4} \left[ \left( \frac{215}{48} - \frac{15}{32} e^2 + \frac{15}{4} \sqrt{1-e^2} \right) - \frac{35 J_4}{18 J_2^2} \left( \frac{21}{4} + \frac{81}{16} e^2 \right) \right]$$

$$B = \frac{3nJ_2 R_e^2}{4p^2} \left\{ -5 + \frac{3 J_2 R_e^2}{p^2} \left[ -\left( \frac{103}{12} + \frac{3}{8} e^2 + \frac{11}{2} \sqrt{1-e^2} \right) + \frac{35 J_4}{6 J_2^2} \left( \frac{31}{14} + \frac{9}{4} e^2 \right) \right] \right\}$$

$$C = \frac{3nJ_2 R_e^2}{4p^2} \left\{ 4 + \frac{3 J_2 R_e^2}{p^2} \left[ \left( 4 + \frac{7}{12} e^2 + 2\sqrt{1-e^2} \right) - \frac{5 J_4}{6 J_2^2} \left( 4 + \frac{9}{2} e^2 \right) \right] \right\}$$

and $p$ is semiparameter, where $p = a(1-e^2)$. It is evident that $f(\sin i)$ is a quartic equation of $\sin i$, so Eq. (11) may have none, two or four real roots for the inclination $i$ in the interval $[-\pi/2, \pi/2]$. Because of the symmetry of Eq. (11), the critical inclination may has none, one or two solutions in the interval $[0, \pi/2]$. The discriminant of Eq. (11) can be expressed as

$$\Delta = B^2 - 4AC \tag{12}$$

If $\Delta < 0$, there exists no critical inclination; if $\Delta \geq 0$, the case is a little more complicated. It can be seen that

$$f(0) = C \tag{13}$$

$$f(1) = A + B + C \tag{14}$$

It is assumed that $J_2$ is positive and $J_4$ is negative in this paper, so both coefficients $A$ and $C$ are positive for any values of the involved parameters. Therefore, $f(0) = C$ is all time positive. Then, Eq. (11) would have two real roots in the interval $[0, \pi/2]$ if $\Delta > 0$, $f(1) \geq 0$ and $0 < -\frac{B}{2A} < 1$, and would have exactly one root in the interval $[0, \pi/2]$ if one of the following conditions are satisfied: (1) $\Delta = 0$, $f(1) \geq 0$, and $0 \leq -\frac{B}{2A} \leq 1$; (2) $\Delta > 0$, and $f(1) < 0$. Otherwise, Eq. (11) has no roots in the interval $[0, \pi/2]$, so the critical inclination does not exist in this case.

## 4.1 The relationship between the multiplicity of the solutions for the critical inclination and the values of $J_2$ and $J_4$

In order to show the relationship between the multiplicity of the solutions for the critical inclination and the terms of $J_2$ and $J_4$, different values of $J_2$ and $J_4$ are assumed in order to obtain general results.

Given the semimajor axis $a=2$, and eccentricity $e=0.1$, different conditions are calculated for a wide range of combinations of $J_2$ and $J_4$. Figure 2 shows the values of the discriminant $\Delta$ as a function of $J_2$ and $J_4$. It can be seen that the discriminant $\Delta$ is always positive at the range of $J_2 \in \left[1\times 10^{-6}, 1\times 10^{-3}\right]$ and $J_4 \in \left[-1\times 10^{-3}, -1\times 10^{-6}\right]$.

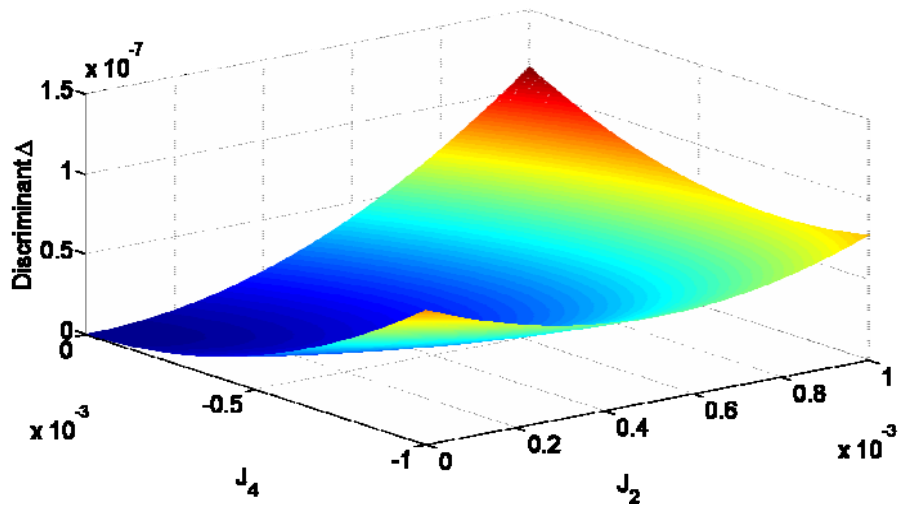

**Fig. 2** The discriminant $\Delta$ as a function of $J_2$ and $J_4$ at the range of $J_2 \in \left[1\times 10^{-6}, 1\times 10^{-3}\right]$ and $J_4 \in \left[-1\times 10^{-3}, -1\times 10^{-6}\right]$ for $a=2$ and $e=0.1$.

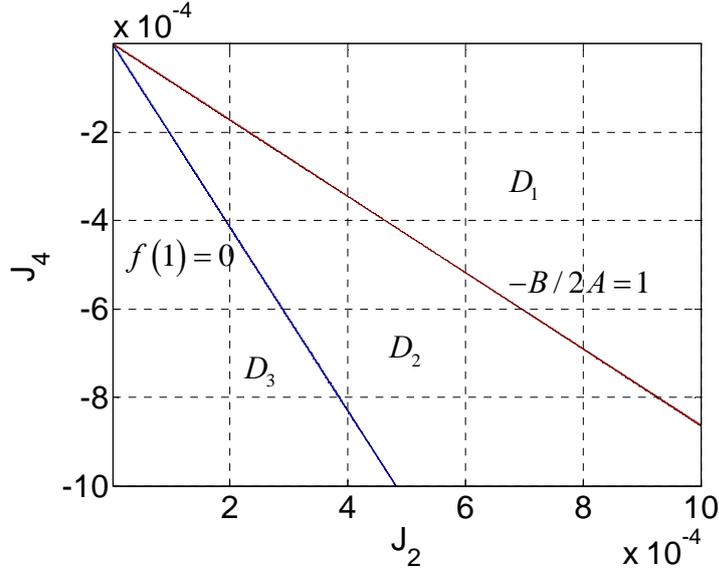

(**a**)

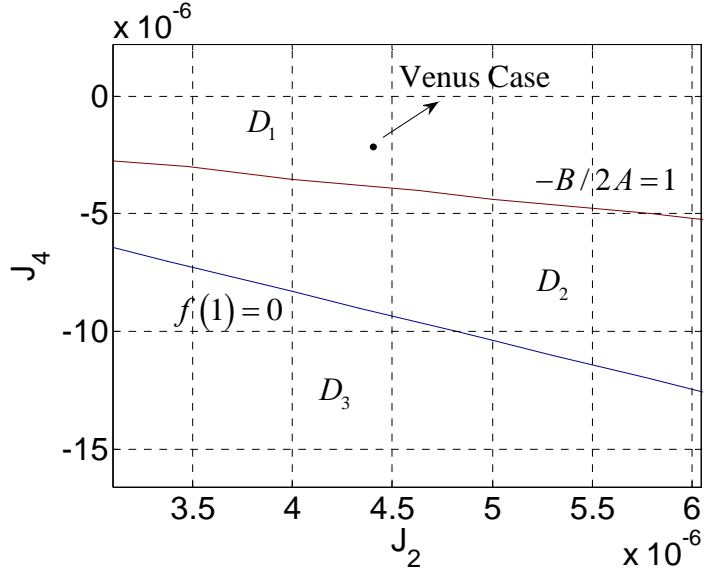

(**b**)

**Fig. 3** (**a**) The contours of $f(1)=0$ and $-B/2A=1$ at the range of $J_2 \in \left[1\times10^{-6}, 1\times10^{-3}\right]$ and $J_4 \in \left[-1\times10^{-3}, -1\times10^{-6}\right]$ for $a$=2 and $e$=0.1. (**b**) Zoom of a part of Fig. 3(a), showing the area in the vicinity of the Venus case.

Figure 3(a) shows the contours of $f(1)=0$ and $-B/2A$=1 at the range of $J_2 \in \left[1\times10^{-6}, 1\times10^{-3}\right]$ and $J_4 \in \left[-1\times10^{-3}, -1\times10^{-6}\right]$. The lines $f(1)=0$ and $-B/2A=1$ divide the plane into three regions: $D_1$, $D_2$ and $D_3$. In the regions $D_1$ and $D_2$, $f(1)<0$, so there exists only one solution for the critical inclination. In the region $D_3$, $f(1)>0$ and $0<-B/2A<1$, so there exist two solutions for the critical inclination in the interval $[0, \pi/2]$. Figure 3(b) is a zoom of Fig. 3(a), and shows that the Venus case in the region $D_1$. The simulation of the Venus case is presented in Fig. 1(a), and it can be seen the drift rates of $e$ and $\omega$ are approximately equal to zero. A combination of $J_2=1\times 10^{-4}$ and $J_4=-3\times 10^{-4}$ is chosen in $D_3$, and the corresponding critical inclinations can be calculated as: $i_{c1}=45.9285°$ and $i_{c2}=78.6215°$. Their simulations in the $J_2$ and $J_4$ gravity field are presented in Fig. 4(a) and 4(b), and show that excursions in $e$ and $\omega$ maintain in the local zones of the initial values.

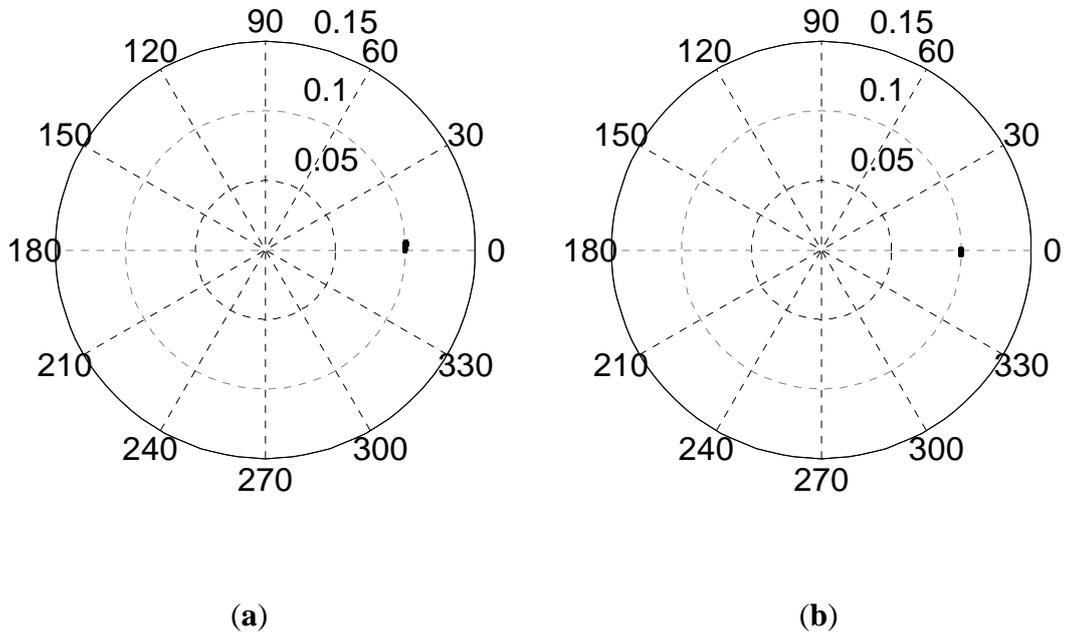

(**a**)          (**b**)

**Fig. 4** $e - \omega$ evolution in the $J_2$ and $J_4$ gravity field over 100 days. (**a**) $i_{c1}=45.9285°$. (**b**) $i_{c2}=78.6215°$.

The cases of $a=1.5$ and $e=0.1$, $a=1.2$ and $e=0.1$, $a=1.5$ and $e=0.2$, $a=1.5$ and

$e$=0.3 are shown in Figs. 5, 6, 7, and 8, respectively. It is found that for these cases, the values of the discriminant $\Delta$ are all positive at the range of $J_2 \in \left[1\times10^{-6}, 1\times10^{-3}\right]$ and $J_4 \in \left[-1\times10^{-3}, -1\times10^{-6}\right]$, and there also exist regions $D_1$ and $D_2$ for one critical inclination and the region $D_3$ for two critical inclinations in the interval $\left[0, \pi/2\right]$. Comparing Figs. 3(a), 5(b) and 6(b), it can be seen that with the decrease of semimajor axis, the range of $J_2$ and $J_4$ for two solutions of the critical inclination becomes larger. While comparing Figs. 5(b), 7(b) and 8(b), it shows that with the increase of eccentricity, the range of $J_2$ and $J_4$ for two solutions of the critical inclination becomes larger.

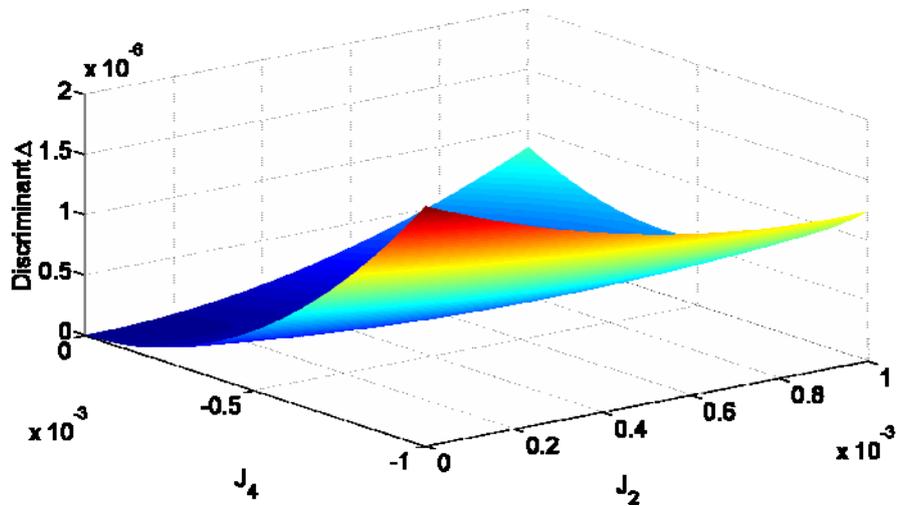

(**a**)

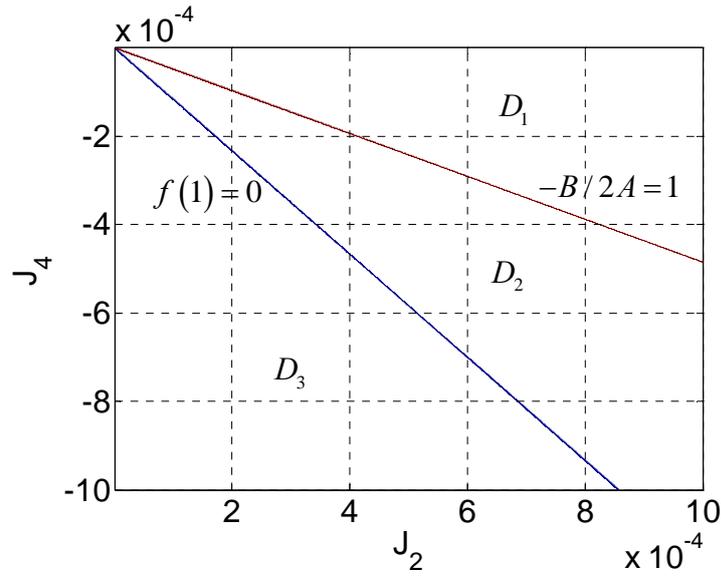

(**b**)

**Fig. 5** The values of the discriminant $\Delta$, $f(0)$, and contours of $f(1)=0$ and $-B/2A=1$ at the range of $J_2 \in \left[1\times10^{-6}, 1\times10^{-3}\right]$ and $J_4 \in \left[-1\times10^{-3}, -1\times10^{-6}\right]$ for $a$=1.5 and $e$=0.1. (**a**) The discriminant $\Delta$. (**b**) The contours of $f(1)=0$ and $-B/2A=1$.

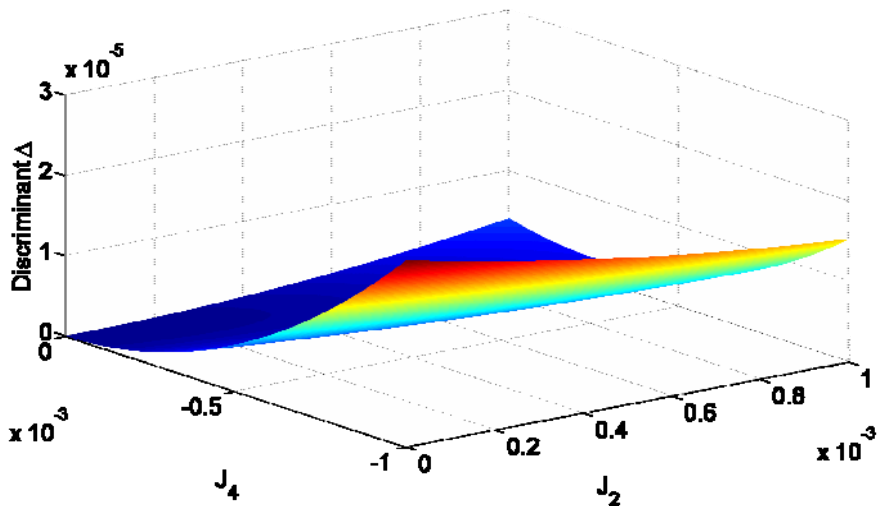

(**a**)

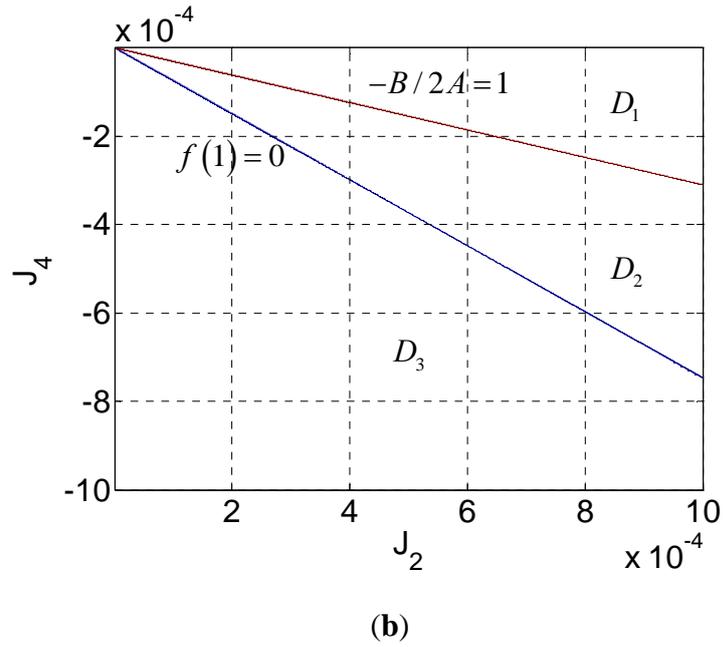

(b)

**Fig. 6** The values of the discriminant $\Delta$, $f(0)$, and contours of $f(1)=0$ and $-B/2A=1$ at the range of $J_2 \in \left[1\times10^{-6}, 1\times10^{-3}\right]$ and $J_4 \in \left[-1\times10^{-3}, -1\times10^{-6}\right]$ for $a$=1.2 and $e$=0.1. (**a**) The discriminant $\Delta$. (**b**) The contours of $f(1)=0$ and $-B/2A=1$.

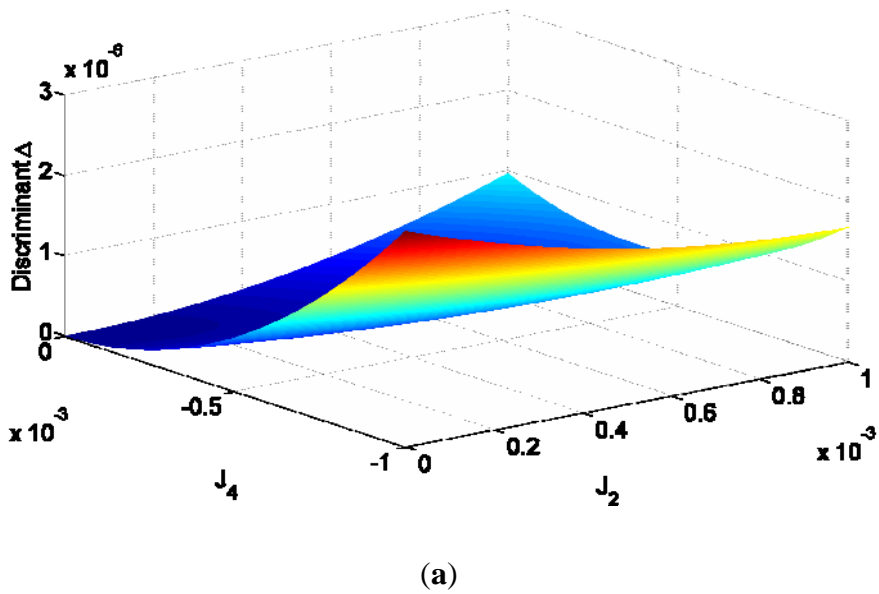

(a)

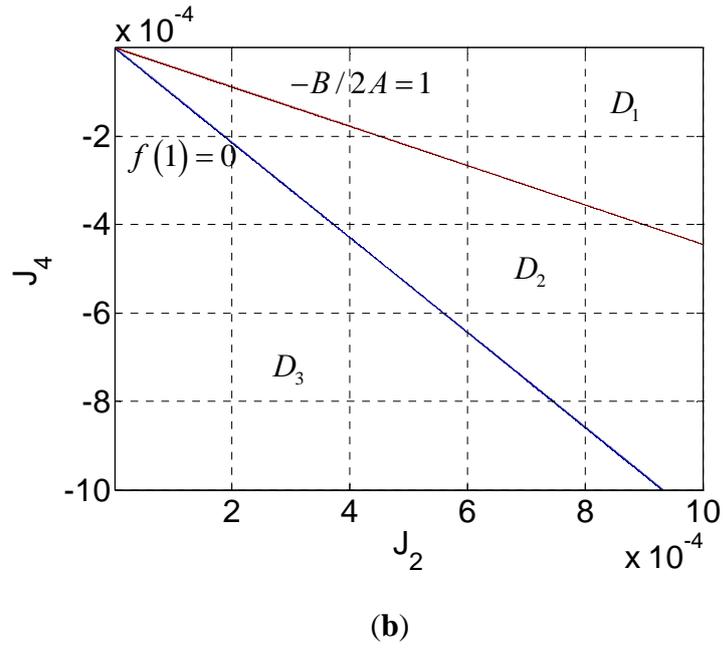

(b)

**Fig. 7** The values of the discriminant $\Delta$, $f(0)$, and contours of $f(1)=0$ and $-B/2A=1$ at the range of $J_2 \in \left[1\times 10^{-6}, 1\times 10^{-3}\right]$ and $J_4 \in \left[-1\times 10^{-3}, -1\times 10^{-6}\right]$ for $a=1.5$ and $e=0.2$. (**a**) The discriminant $\Delta$. (**b**) The contours of $f(1)=0$ and $-B/2A=1$.

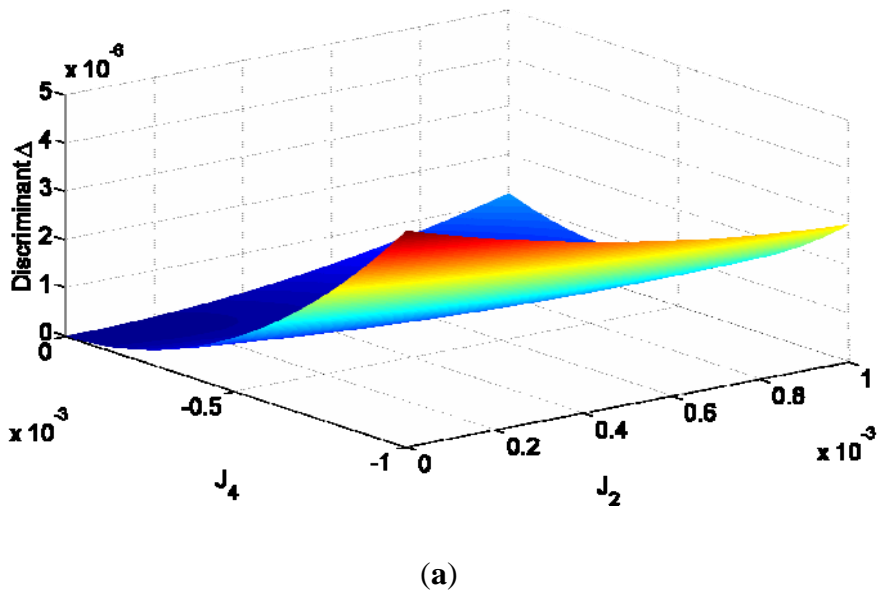

(a)

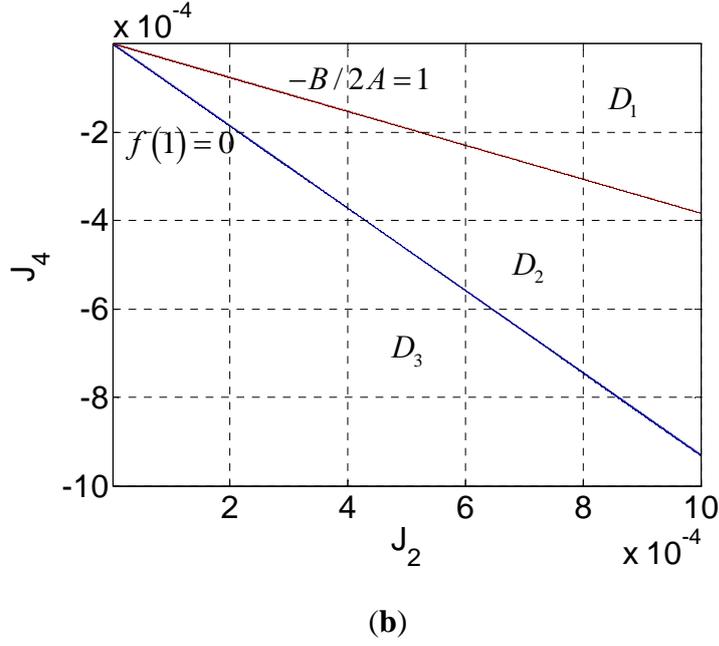

(b)

**Fig. 8** The values of the discriminant $\Delta$, $f(0)$, and contours of $f(1)=0$ and $-B/2A=1$ at the range of $J_2 \in \left[1\times10^{-6}, 1\times10^{-3}\right]$ and $J_4 \in \left[-1\times10^{-3}, -1\times10^{-6}\right]$ for $a=1.5$ and $e=0.3$. (**a**) The discriminant $\Delta$. (**b**) The contours of $f(1)=0$ and $-B/2A=1$.

## 4.2 The relationship between the multiplicity of the solutions for the critical inclination and the values of semimajor axis and eccentricity

When the values of $J_2$ and $J_4$ are fixed, the multiplicity of the solutions for the critical inclination depends on the values of semimajor axis and eccentricity. The values of $J_2=4\times10^{-6}$ and $J_4=-5\times10^{-6}$ is selected, where $J_2$ and $J_4$ are of the same order of magnitude as Venusian $J_2$ and $J_4$, respectively. Then different conditions are calculated for a wide range of combinations of semimajor axis and eccentricity. The values of the discriminant $\Delta$ as a function of semimajor axis and eccentricity at the range of $a \in [1,3]$ and $e \in [0, 1-R_e/a]$ are shown in Fig. 9. The upper bound of eccentricity is set equal to $1-R_e/a$ in order to avoid collision with the central body. It can be seen that the discriminant is always positive. Figure 10 presents the variations of the critical inclination as a function of $a$ and $e$ at the range of $a \in [1, 3]$ and $e \in [0, 1 - R_e/a]$ for fixed $J_2$ and $J_4$. It can be seen that the critical inclination may has one or two solutions.

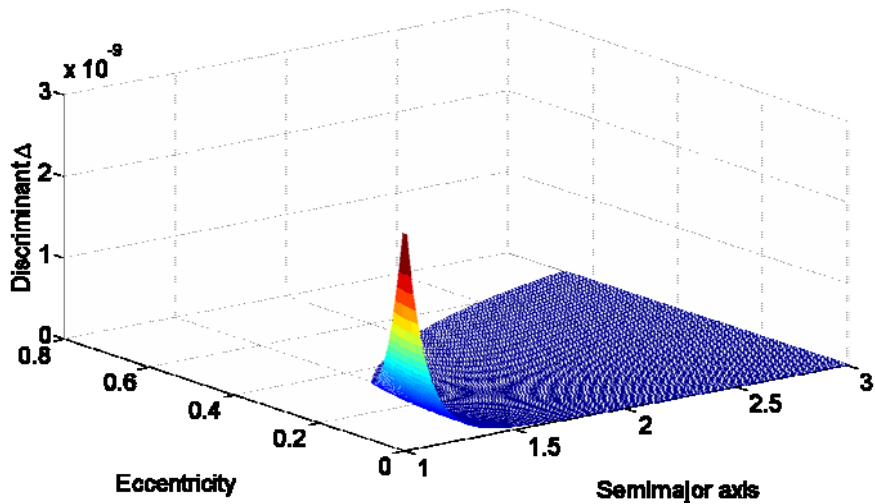

**Fig. 9** The discriminant $\Delta$ as a function of semimajor axis and eccentricity at the range of $a \in [1, 3]$ and $e \in [0, 1 - R_e/a]$.

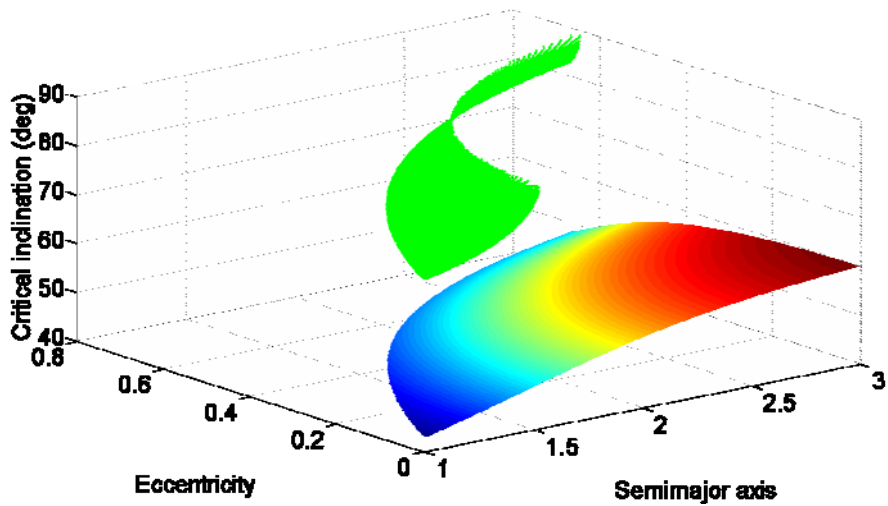

**Fig. 10** The variations of the critical inclination as a function of semimajor axis and eccentricity at the range of $a \in [1, 3]$ and $e \in [0, 1 - R_e / a]$.

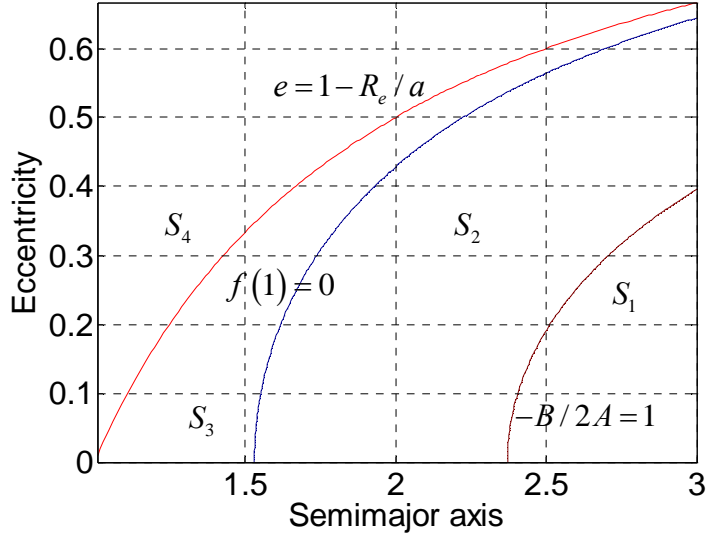

**Fig. 11** The contours of $f(1) = 0$ and $-B/2A = 1$ at the range of $a \in [1, 3]$ and $e \in [0, 1 - R_e / a]$.

Figure 11 shows the contours of $f(1) = 0$ and $-B/2A=1$ at the range of $a \in [1, 3]$ and $e \in [0, 1 - R_e / a]$. The lines $f(1) = 0$ and $-B/2A=1$ divide the plane into four regions: $S_1$, $S_2$, $S_3$ and $S_4$. In the regions $S_1$ and $S_2$, $f(1) < 0$, so there exists only one solution for the critical inclination. In the region $S_3$, $f(1) > 0$ and $0 < -B/2A < 1$, so there exist two solutions for the critical inclination in the interval $[0, \pi/2]$. In the region $S_4$, $e > 1 - R_e / a$, so this is a crash case. An initial condition in the region $S_2$ is chosen: $a=2$, $e=0.2$, and the corresponding critical inclination can be calculated as: $i_c = 53.2518°$. The simulation of this initial condition in the $J_2$ and $J_4$ gravity field is presented in Fig. 12, and it shows a good frozen property. For the initial condition $a=1.5$, $e=0.2$ in the region $S_3$, the corresponding critical inclinations can be calculated as: $i_{c1}=47.7279°$ and $i_{c2}=82.0926°$, and their simulations in the $J_2$ and $J_4$ gravity field are presented in Fig. 13(a) and 13(b). They also show good frozen properties.

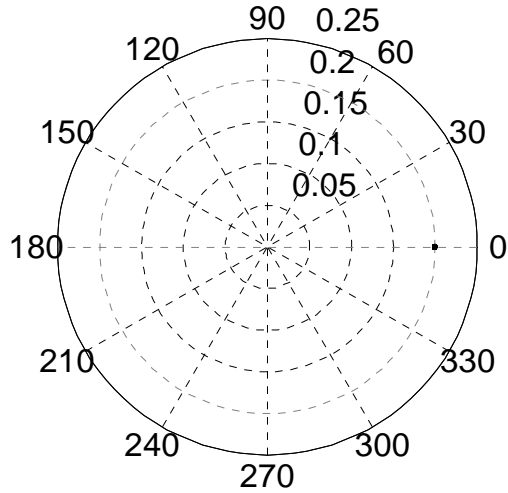

**Fig. 12** $e - \omega$ evolution in the $J_2$ and $J_4$ gravity field over 100 days for $i_c=53.2518°$.

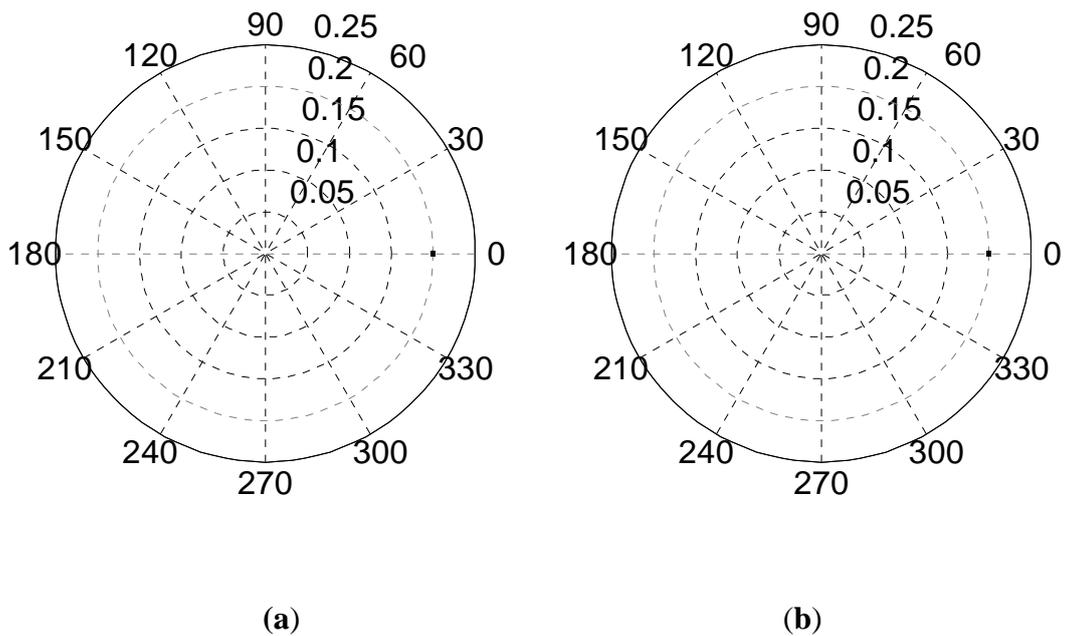

(a)  (b)

**Fig. 13** $e - \omega$ evolution in the $J_2$ and $J_4$ gravity field over 100 days. (**a**) $i_{c1}$=47.7279°. (**b**) $i_{c2}$=82.0926°.

## 5. Conclusions

This paper investigates the extended problem of the critical inclination, and obtains some useful results. For celestial bodies where the oblateness term $J_2$ is not dominant among the harmonic coefficients, the concept of the critical inclination is different from that of the traditional sense. When given the values of argument of pericentre and eccentricity, the relationship between the multiplicity of the solutions for the critical inclination and the values of $J_2$ and $J_4$ is analyzed. The ranges of $J_2$ and $J_4$ for one critical inclination far away from that of the traditional sense and for two critical inclinations in the interval $[0, \pi/2]$ are found. It shows that the Venusian $J_2$ and $J_4$ lie on the region of one solution for the critical inclination. Besides, when given certain values of $J_2$ and $J_4$, the relationship between the multiplicity of the solutions for the critical inclination and the values of semimajor axis and eccentricity is studied. The ranges of semimajor axis and eccentricity for the crash case, for one critical inclination far away from that of the traditional sense, and for two critical inclinations in the interval $[0, \pi/2]$ are obtained. This study shows that for some cases, the value of the critical inclination is far away from that of the traditional sense, or even has multiple solutions.

**Acknowledgments**

This work was supported by the National Natural Science Foundation of China (No. 10832004 and No. 11072122).

**References**

Allan, R. R.: The critical inclination problem: A simple treatment. Celest. Mech. Dyn. Astron. **2**(1), 121–122 (1970). doi: 10.1007/BF01230456

Aorpimai, M., Palmer P. L.: Analysis of frozen conditions and optimal frozen orbit insertion. J. Guid. Control. Dyn. **26**(5), 786–793 (2003). doi: 10.2514/2.5113

Barker, L., Stoen, J.: Sirius satellite design: The challenges of the Tundra orbit in commercial spacecraft design. Adv. Astronaut. Sci. **107**, 575–596 (2001)

Brouwer, D.: Outlines of general theories of the Hill-Brown and Delaunay types for orbits of artificial satellites. Astron. J. **63**, 433–438 (1958)

Brouwer, D.: Solution of the problem of artificial satellite theory without drag. Astron. J. **64**(1274), 378–397 (1959). doi: 10.1086/107958

Bruno, M. J., Pernicka, H. J.: Mission design considerations for the Tundra constellation. AIAA/AAS Astrodynamics Specialist Conference and Exhibit, AIAA 2002–4634 (2002)

Bruno, M. J., Pernicka, H. J.: Tundra constellation design and stationkeeping. J. Spacecr. Rocket. **42**(5), 902–912 (2005). doi: 10.2514/1.7765

Chobotov, V.A.: Orbital Mechanics. AIAA, Reston (2002)

Coffey, S. L., Deprit, A., Deprit, E.: Frozen orbits for satellites close to an Earth-like planet. Celest. Mech. Dyn. Astron. **59**(1), 37–72 (1994). doi: 10.1007/BF00691970

Coffey, S. L., Deprit A., Miller, B. R.: The critical inclination in artificial satellite theory. Celest. Mech. Dyn. Astron. **39**(4), 365–406 (1986). doi: 10.1007/BF01230483

Delhaise, F., Morbidelli, A.: Luni-solar effects of geosynchronous orbits at the critical inclination. Celest. Mech. Dyn. Astron. **57**(1–2), 155–173 (1993)

Ely, T. A., Lieb, E.: Constellations of elliptical inclined lunar orbits providing polar and global coverage. AAS/AIAA Astrodynamics Specialist Conference, AAS 05–343 (2005)

Garfinkel, B.: On the motion of a satellite in the vicinity of the critical inclination. Astron. J. **65**(10), 624–627 (1960). doi: 10.1086/108308

Garfinkel, B.: The global solution of the problem of the critical inclination. Celest. Mech. Dyn. Astron. **8**(1), 25–44 (1973). doi: 10.1007/BF01228388

Gunning G. R. Chao C. C.: A PC based tool for Molniya orbit analysis. AIAA/AAS Astrodynamics Conference, AIAA 96–3657 (1996)

Izsak, G. I.: On the critical inclination in satellite theory. SAO Special Report #90. 1962

Jupp, A. H.: The critical inclination problem – 30 years of process. Celest. Mech. Dyn. Astron. **43**(1–4), 127–138, (1988) doi: 10.1007/BF01234560

Kidder, S. Q., Vonder Haar, T. H.: On the use of satellites in Molniya orbits for meteorological observation of middle and high latitudes. J. Atmos. Ocean. Tech. **7**, 517–522 (1990)

Konopliv, A. S., Asmar, S. W., Carranza, E., Sjogren, W. L., Yuan, D. N.: Recent gravity models as a result of the Lunar Prospector mission. Icarus. **150**, 1–18( 2001)

Konopliv, A. S., Banerdt, W. B., Sjogren, W. L.: Venus gravity: 180th degree and order model. Icarus. **139**, 3–18 (1999)

Kozai, Y.: The motion of a close Earth satellite. Astron. J. **64**(1274), 367–377 (1959). doi: 10.1086/107957

Lara, M., Deprit, A., Elipe, A.: numerical continuation of families of frozen orbits in the zonal problem of artificial satellite theory. Celest. Mech. Dyn. Astron. **62**(2), 167–181 (1995). doi: 007/BF00692085

Lemoine, F. G., Kenyon, S. C., Factor, J. K., Trimmer, R.G., Pavlis, N. K., Chinn, D. S., Cox, C. M., Klosko, S. M., Luthcke, S. B., Torrence, M. H.,Wang, Y. M.,Williamson, R. G., Pavlis, E. C., Rapp, R. H., Olsonet, T. R.: The development of the joint NASA GSFC and NIMA geopotential model EGM96. NASA TM 1998–206861, 1998

Lemoine, F. G., Smith, D. E., Rowlands, D. D., Zuber, M. T., Neumann, G. A., Chinn, D. S., Pavlis, D. E.: An improved solution of the gravity field of Mars (GMM-2B) from Mars Global Surveyor. J. Geophys. Res. **106**(E10), 23,359–23,376 (2001) doi:10.1029/2000JE001426

Liu, X., Baoyin, H., Ma, X.: Analytical investigations of quasi-circular frozen orbits in the Martian gravity field. Celest. Mech. Dyn. Astron. (2011) doi: 10.1007/s10569-010-9330-2

Orlov, A. A.: Pochti Krugovye Periodicheskie Dvizheniia Materialnoi Tochki Pod Deistviem Niutonovskogo Pritiazheniia Sferoida. Soobshcheniia Gosudarstvennogo astronomicheskogo instituta imeni P.K. Shternberga. (88–89), 3–38 (1953)

Russell, R. P., Lara, M.: Long-lifetime lunar repeat ground track orbits. J. Guid. Control. Dyn. **30**(4), 982–993 (2007)

Saedeleer, B. De., Henrard, J.: The combined effect of $J_2$ and $C_{22}$ on the critical inclination of a lunar orbiter. Adv. Space. Res. **37**(1), 80–87 (2006). doi: 10.1016/j.asr.2005.06.052

Stone, A. D., Brodsky, R. F.: Molniya orbits obtained by the two-Burn method. J. Guid. Control. Dyn. **11**(4), 372–375 (1988)